# Multiscale insight into the $Cd_{1-x}Zn_xTe$ vibrational-mechanical properties – High-pressure experiments and *ab initio* calculations

T. Alhaddad,[1] M.B. Shoker,[2,a)] O. Pagès,[1,b)] A. Polian,[3] V.J.B. Torres,[4] Y. Le Godec,[5] J.-P. Itié,[5] C. Bellin,[3] K. Béneut,[3] S. Diliberto,[6] S. Michel,[6] A. Marasek,[7] and K. Strzałkowski[7]

[1] Université de Lorraine, LCP-A2MC, ER 4632, F-57000 Metz, France
[2] University of Luxembourg, Department of physics and materials science, 41 rue du Brill, 4422 Belvaux, Luxembourg
[3] Institut de Minéralogie, de Physique des Matériaux et de Cosmochimie, Sorbonne Université — UMR CNRS 7590, F-75005 Paris, France
[4] Departamento de Fisica and I3N, Universidade de Aveiro, 3810 – 193 Aveiro, Portugal
[5] Synchrotron SOLEIL, L'Orme des Merisiers Saint-Aubin, BP 48 F-91192 Gif-sur-Yvette Cedex, France
[6] Institut Jean Lamour, UMR CNRS 7198, Campus Artem, Université de Lorraine, 57078, Metz, France
[7] Institute of Physics, Faculty of Physics, Astronomy and Informatics, Nicolaus Copernicus University in Toruń, ul. Grudziądzka 5, 87-100 Toruń, Poland

**Abstract.** The $Cd_{1-x}Zn_xTe$ semiconductor alloy is a regular system regarding its macroscopic mechanic properties in that its experimental bulk modulus exhibits a linear $x$-dependence, in line with *ab initio* predictions. Complexity arises at the bond scale, referring to the intricate $Cd_{1-x}Zn_xTe$ percolation-type Raman pattern [T. Alhaddad *et al.*, J. App. Phys. **133**, 065701 (2023)]. This offers an appealing benchmark to test various phonon coupling processes at diverse length scales in a compact multi-oscillator assembly, presently tuned by pressure. At $x{\sim}0$, an inter-bond long-range/macro electric coupling between the matrix and impurity polar phonons is detuned under pressure. Inversely, at $x{\sim}1$, an intra-bond short-range/nano mechanic coupling is enforced between the two Zn-Te apolar sub-phonons stemming from "same" and "alien" percolation-type environments. The pressure-induced macro/nano polar/apolar coupling/decoupling processes are compared within a model of two coupled electric/mechanic harmonic oscillators in terms of a compromise between proximity to resonance and strength of coupling, impacting the degree of mode mixing, with *ab initio* (apolar case) and analytical (polar case) Raman calculations in support. Notably, the free mechanic coupling at $x{\sim}1$ opposes the achievement of a phonon exceptional point, manifesting the inhibition of mechanic coupling, earlier evidenced with similar bonds for $x \leq 0.5$. Hence, the pressure dependence of a given bond vibration in a disordered alloy basically differs depending on whether the bond is matrix-like, *i.e.*, self-connected in bulk (free coupling), or dispersed, *i.e.*, self-connected in a chain (inhibited coupling). This features pressure-tunable percolation-based on-off phonon switches in complex media.

---

a) This research was performed while M.B. Shoker was doing his PhD at Laboratory LCP-A2MC (affiliation 1).
b) Author to whom correspondence and requests for materials should be addressed: olivier.pages@univ-lorraine.fr

**Introduction.** III-V and II-VI $A_{1-x}B_xC$ zincblende ternary semiconductor alloys, made of merely two bond species – against three in the IV $A_{1-x}B_x$-binaries with diamond structure – randomly arranged (in the ideal case) on a regular lattice with highest (cubic) symmetry, form ultimately simple systems out of complex media.[1] In fact, they assimilate with geometrical composite objects, setting a benchmark for studying how the physical properties are impacted by disorder in a context of percolation.[2] For doing so, one further needs a sensitive probe at the atom scale where the A↔B substitution operates. A suitable candidate is the effective bond force constant, conveniently accessed by Raman scattering.

A unified understanding, hitherto missing, of the Raman spectra of cubic alloys, covering the III-V and II-VI ternaries along with the IV binaries, has lately been achieved within our so-called percolation model (PM).[3] In brief, the PM formalizes a sensitivity of bond vibrations to their "same" or "alien" environments defined at the nanoscale, *i.e.*, up to first or at most second neighbors, or with a mix of the two. This is consistent with a basic rule that phonons are essentially a matter of short-range interactions.[4] Such description of a disordered alloy in terms of a nanostructured composite introduces a bimodal Raman signal per bond in II-VI and III-V ternaries, further diversifying into a trimodal pattern in IV binaries.[5] The multi-mode fine structure shows up distinctly only for the short/stiff/light bond vibrating at high frequency that mostly accommodates the mismatch in bond physical properties.[3,5] As such, the PM markedly deviates from the historical modified-random-element-isodisplacement (MREI) model[6] that views an alloy as a uniform continuum of the same virtual $A_{1-x}B_x$ atom (intercalated with the invariant C species in III-V and II-VI ternaries). By construction, such virtual crystal approximation skips the mesostructure of a disordered alloy, simplifying the Raman signal to a unique mode per bond.

Our unified Raman understanding of cubic-alloys at ambient pressure within the PM was lately completed by a predictive PM-based bipartition of II-VI and III-V ternaries depending on whether the percolation-type Raman fine structure converges or diverges under pressure,[7] as fixed by the relative hardening rates of the "same" and "alien" environments.[8]

Among re-examined alloys within the PM so far, cubic $Cd_{1-x}Zn_xTe$, studied in this work, is outstanding.[3] At $x \sim 0$, its Raman signal fits into the MREI scheme. A remarkable feature is that the upper/minor longitudinal optic (LO) impurity mode ($LO_{Zn-Te}$) and the lower/main matrix one ($LO_{Cd-Te}$) are quasi resonant and couple via their common macroscopic electric field $E$ reflecting the ionicity/polarity of the chemical bonding.[9] This results in a pair of ($LO^-, LO^+$) features in mutual repulsion with similar intensities. At $x \sim 1$, the more refined PM description prevails, manifested by a distinct splitting of the apolar (purely-mechanic, *i.e.*, deprived of electric field[9]) transverse optic (TO) mode due to the dominant Zn-Te bond into two lower/main $TO_{Zn-Te}^{Zn}$ and upper/minor $TO_{Zn-Te}^{Cd}$ uncoupled subTO's stemming from "same" and "alien" environments (specified by a superscript). Altogether, this forms an intricate phonon pattern combining long-range polar-LO coupling with short-range apolar-TO decoupling. Such situations can *a priori* be tuned by pressure, ideally up to on-off phonon switches, as investigated in this work.

The TO case is especially interesting. It gives opportunity to test experimentally how the short-range mechanic coupling develops within the percolation doublet of a matrix-like bond, *i.e.*, self-connected in bulk, when its two submodes are forced to proximity by pressure. Previous experiments on alloys classified of the converging type in our PM-bipartition,[7] as opposed to the diverging type, pertained to dispersed bonds, *i.e.*, self-connected in a chain. A phonon exceptional point was then observed at the resonance, manifesting the inhibition of mechanic coupling.[7] In this scenario 1, only the sub-TO due to the "alien" environment survives the resonance; that stemming from the "same" environment freezes.[7] It does not make sense that scenario 1 applies to a matrix-like bond. Taken to extreme, this would mean the "freezing" of a crystal in bulk as soon as corrupted by an impurity. Preliminary *ab initio* tests on BeTe-like supercells containing a unique Cd or Zn impurity,[10,11] sufficient to distinguish between "same" (away from impurity) and "alien" (close to impurity) environments of a matrix-like bond, revealed the doublet inversion under pressure. Hence, both subTO's would survive the resonance and the exceptional point would be avoided. This suggests a free mechanic coupling across the resonance. However, such scenario 2 was observed at extreme impurity dilution only, and is clearly attested only in *ab initio* data. Experimental support exists for the long/soft/heavy Zn-Te bond of $Zn_{1-x}Be_xTe$. However, the percolation doublet is hardly discernible.[7]

A decisive test should rely on experiment done on the short/stiff/light bond, with a number of requirements on the used $A_{1-x}B_xC$ alloy. First, the $A \leftrightarrow B$ substitution should be random,[12] to ascertain that the observed phenomena are intrinsic to alloying. Second, the crystal should be available in bulk,[13] as needed to perform high-pressure experiments. Last, the PM-doublet should mark the matrix-like bond, and, moreover, be of the converging type under pressure.[7] Among re-examined alloys within the PM so far, the only candidate that fulfills all requirements is $Cd_{1-x}Zn_xTe$ taken at large Zn content.[3]

In this work, the pressure dependence of the odd inter-bond macro-LO ($x \sim 0.1$) and intra-bond nano-TO ($x \sim 0.9$) zincblende-$Cd_{1-x}Zn_xTe$ modes is studied by high-pressure Raman scattering (HP-RS) using high quality single crystals grown by the Bridgman method.[13] A similar $Cd_{1-x}Be_xTe$[11] ($x=0.11$) crystal is used for calibration purpose. High-pressure X-ray diffraction (HP-XRD) is further used to access the $Cd_{1-x}Zn_xTe$ bulk modulus at ambient pressure $B_0$. Altogether, the HP-RS and HP-XRD studies, supported by *ab initio* calculations, complete an overview of the $Cd_{1-x}Zn_xTe$ mechanical-vibrational properties at the bulk/macro ($B_0$, LO) and bond/nano (TO) scales throughout the composition domain.

**Methods.** The HP-RS is performed in-house on a single crystal in a Chervin-type diamond anvil cell[14] (DAC, ~350 μm diamond culet diameter) using 16:3:1 Methanol-ethanol-water mixture, hydrostatic up to ~10.5 GPa.[15] Ruby spheres luminescence was used as a pressure gauge.[16] HP-XRD studies are done at the PSICHE beamline of SOLEIL synchrotron on powders using a similar Chervin-type DAC, with Ne as a pressure transmitting medium, hydrostatic up to ~15 GPa.[15] Au powder was added for pressure calibration. The HP-XRDiffractograms, obtained with the 0.3738 Å radiation, are treated via the DIOPTAS software.[17] $B_0$ is derived by fitting a third-order Birch-Murnaghan's equation of state to the pressure ($P$) dependence of the unit cell volume in the native zincblende phase.[18] In HP-RS data, a selective Fröhlich-enhanced[19] LO insight at $x \sim 0.1$ is achieved by using the 488.0 nm Ar$^+$ laser line, near-resonant with the direct $E_0 + \Delta_0$ gap between the lower light hole valence band and the conduction band.[3] Low temperature (77 K) is used to minimize the formation of Te inclusions, a notorious problem with CdTe-like crystals,[20] presently emphasized by laser heating of the tiny absorbing sample inside the DAC. In such crystals, $E_0$ increases with pressure[21] and by decreasing temperature,[22] which improves the resonance, further enhancing the LO's. At $x \sim 0.9$, $Cd_{1-x}Zn_xTe$ is transparent to the used 632.8 nm laser line.[2] This suppresses the need for low temperature and also the LO enhancement, hence clarifying the TO insight.

Contour modeling of the TO and LO Raman spectra is achieved within a dielectric approach[23] via our generic expression of the Raman cross section (RCS) set in Eq. (1) of Ref.[3]. For the TO-coupled case we use an *ad hoc* variant lately derived in Ref.[10], referring to Eq. (7) therein. A sensitive input parameter is the pressure dependence of the parent oscillator strength $S$. This is well-documented only for ZnTe, not for CdTe. A direct insight in the latter case is achieved *in situ* by measuring the TO-LO splitting manifested by the uncoupled Cd-Te mode of $Cd_{0.89}Be_{0.11}Te$ placed nearby $Cd_{0.89}Zn_{0.11}Te$ inside the DAC (Fig. S1; S stands for supplementary material), used as an internal calibrator. Detail is given in Sec. SII.3.

*Ab initio* insights into the $Cd_{1-x}Zn_xTe$ bulk modulus $B_0$ and TO Raman spectra are obtained by applying the AIMPRO (*Ab Initio* Modeling PROgram) code[24,25] to large (216-atom) fully relaxed (lattice constant, atom positions) zincblende-supercells optimized to a random $Cd \leftrightarrow Zn$ substitution. This is achieved by virtual annealing until the distribution of individual Te-centered Cd/Zn-tetrahedra matches the binomial Bernoulli's law.[3] $B_0$ follows from *ab initio* data in the same way as from HP-XRD ones. The TO Raman spectra are calculated using the formula of de Gironcoli.[26]

**Results and discussion.** First, we address the $Cd_{1-x}Zn_xTe$ mechanic properties in bulk via $B_0$ by HP-XRD, further needed to identify the pressure domain of the native zincblende phase next studied by HP-RS.

Not surprisingly, the phase sequences at moderate Zn- and Cd-dilutions replicate the CdTe (cubic-zincblende, hexagonal-cinnabar, cubic-rock-salt, orthorhombic-Cmcm, abbreviated zb→cb→rs→cm) and ZnTe (zb→cb→cm) ones (Fig. S2), respectively.[27] Only, $Cd_{0.72}Zn_{0.28}Te$ transites directly from zb to rs, the transient cb phase being skipped. Regarding the HP-RS studies, Cd↔Zn alloying around Te conveniently enlarges the zincblende domain of the CdTe-like $Cd_{0.90}Zn_{0.1}Te$ (~5.5 GPa)[29] and $Cd_{0.72}Zn_{0.28}Te$ (~6 GPa) alloys compared with CdTe (~2.5 GPa)[27] and also of the ZnTe-like $Cd_{0.11}Zn_{0.89}Te$ (~10.5 GPa) alloy compared with ZnTe (~8.1 GPa).[28] The limit pressure of the native phase is indicated in brackets. This deviates from the linear composition dependence of the same critical pressure in case of Cd↔Zn alloying around Se.[7] The deviation is presumably due to the exceptionally large covalency of the Zn-Te bond among II-VI's, quasi matching that of III-V's.[8,30] Hence, covalent/stiff Zn-Te bonding in CdTe would harden the lattice and delay the departure from the native zb phase. Reversely, minor alloying with the ionic/soft Cd-Te bond would not challenge the natural rigidity of a ZnTe-like lattice.

The $P$-dependence of the zb lattice constant (Fig. S3), used to derive the volume vs. $P$ variation (Fig. S4) from which $B_0$ is issued, reveals a linear $B_0$ vs. $x$ dependence within experimental error (Fig. 1, filled symbols), aligned with existing parent values measured by HP-XRD.[31,32] The linear trend is nicely echoed in *ab initio* data (Fig. 1, hollow symbols). It was also evidenced with $ZnSe_xTe_{1-x}$ after a thorough HP-XRD study.[31] Hence, $Cd_{1-x}Zn_xTe$ looks "regular" regarding its mechanic behavior in bulk.

From now on the focus shifts to the bond scale addressed by HP-RS. Generally, the LO's and TO's drift upwards under pressure because the Raman frequency squared scales as the effective bond force constant[6] and the Cd-Te and Zn-Te bonds strengthen when the lattice downsizes.[8] Comparing the volume derivatives of the bond ionicities $df_i/dlnV$,[8] Zn-Te hardens at a faster rate than Cd-Te. Hence, the Raman drift should be larger for the Zn-Te modes than for the Cd-Te ones. In this line, the lower CdTe-like $LO^-$ and upper ZnTe-like $LO^+$ should break away under pressure and $\vec{E}$-decouple ($x$~0.1). Oppositely, the lower $TO_{Zn-Te}^{Zn}$ and upper $TO_{Zn-Te}^{Cd}$ should get closer and mechanically couple ($x$~0.9).

At $x$~0.1 the $Cd_{1-x}Zn_xTe$ LO Raman spectrum taken at ambient conditions on a large crystal exhibits distinct ($LO^-, LO^+$) features with comparable intensities around 170 cm$^{-1}$ (Fig. 2a, left spectrum).[3] The same two modes stemming from a tiny sample exposed to minimal pressure in a DAC are blurred (Fig. S5). The blurring persists across the series of $Cd_{0.9}Zn_{0.1}Te$ HP-RS spectra reported by Saqib et al.[29] This might be due to the formation of spurious Te inclusions at the laser spot, evidenced by a strong Raman peak at ~115 cm$^{-1}$. By working at 77 K the Te inclusions are avoided in our case,[33] so that the ($LO^-$, $LO^+$) signal clarifies from 2 GPa onwards (Fig. 2a, right spectrum; Fig. S5).

As expected, $LO^-$ and $LO^+$ diverge under pressure and at the same time $LO^-$ strengthens against $LO^+$ (Fig. 2a), two signs of $\vec{E}$-decoupling. Fair contour modeling of the $Cd_{0.89}Zn_{0.11}Te$ LO Raman signal vs. $P$, covering frequencies (clear curves) and intensities (thickness of curves), is achieved via the LO version of our RCS.[3] To keep physical meaning only one parameter is adjusted (to model 14 peaks in total, varying in frequency and intensity), i.e., the slope of the $TO_{Zn-Te}$ frequency vs. $P$ variation (upper-thick line) – not visible experimentally, taken linear for simplicity. Detail is given in Sec. SII.3a.

At $x$~0.9, the minor-$TO_{Zn-Te}^{Cd}$ and main-$TO_{Zn-Te}^{Zn}$ converge in the upstroke (Fig. 2b), as expected. The two modes cross each other into a unique/symmetric Raman peak at the resonance around 4 GPa (central spectrum) and complete an inversion at ~9 GPa (Fig. S6a). This is attested by a reversal in asymmetry of the overall Zn-Te signal at extreme pressures. The inversion is reversible in the downstroke (Fig. S6a, dotted curve) and replicates with $Cd_{0.13}Zn_{0.87}Te$ (Fig. S6b), hence robust. A fair step-by-step contour modeling of the $Cd_{0.11}Zn_{0.89}Te$ TO Raman signal in course of the inversion, covering frequencies (symbols and clear curves – Fig. 2b; theoretical spectra – Fig. S5) and intensities (thickness of curves – Fig. 2b), is achieved by using a simplified version of our RCS for two mechanically coupled ($TO^-,TO^+$) modes.[10] A weak mechanic coupling is generally assumed, with characteristic frequency $\omega'$~17 cm$^{-1}$ representing ~10% of the ZnTe TO frequency. Nevertheless, $\omega'$ is doubled near the resonance (upper/thin curve – Fig. 2b) where the coupling achieves maximum, to account for the observed large $TO^- - TO^+$ repulsion therein, i.e., ~8 cm$^{-1}$ (marked by opposite arrows). The $P$-dependencies of the raw-uncoupled minor-$TO_{Zn-Te}^{Cd}$ and main-$TO_{Zn-Te}^{Zn}$ frequencies behind ($TO^-,TO^+$) – not shown for clarity – are taken linear between extreme values, for simplicity, and set by analogy with ZnTe,[28] respectively. Detail is given in Sec. SII.3b.

The $P$-induced $TO^{Cd}_{Zn-Te} \leftrightarrow TO^{Zn}_{Zn-Te}$ inversion reveals that both sub-TO's due to a matrix-like bond survive the resonance, at variance with the achievement of a phonon exceptional point earlier evidenced in a similar context with minor-to-moderate bonds.[34] This points towards the bond percolation threshold of the "alien" species, *i.e.*, $x_{Cd-Te}$=0.81, as the pivotal composition separating the free-coupling and exceptional-point regimes of Zn-Te. By definition, $x_{Cd-Te}$ marks the onset of Cd-Te self-connection into an infinite path across the crystal, a purely statistical effect of the random Cd↔Zn substitution.[3] Waiting for experiment, the pivotal role of $x_{Cd-Te}$ is attested in our *ab initio* data showing the $P$-induced emergence (Fig. S7) of a minor Raman feature, assigned as the inverted minor-$TO^{Cd}_{Zn-Te}$, on the low frequency tail of the main-$TO^{Zn}_{Zn-Te}$ only at $x$ close to $x_{Cd-Te}$ ($x$=0.83, 0.79), not far from it ($x$=0.7, 0.5). However, a more intensive *ab initio* test run on large supercells naturally involving the mesostructure of a disordered alloy in its full variety together with related percolative effects, would be needed to ascertain the presumed pivotal role of $x_{Cd-Te}$.

The net efficiencies of the inter-bond LO-electric decoupling at $x\sim0.1$ and intra-bond TO-mechanic coupling at $x\sim0.9$ are compared within a model of two coupled harmonic oscillators (mass+restoring electric/mechanic force) in terms of a compromise between the strength of coupling and the proximity to resonance, given by the off-diagonal and diagonal terms of the dynamic matrix, respectively. The current approach is in line with that used in recent work on overdamped-TO's and LO's.[7] However, presently the TO's are undamped and the LO's are studied vs. $P$, and not vs. $x$. The amount of mode mixing depending on pressure is derived from the coordinates of the unit orthogonal wavevectors of the coupled modes. A large LO-mixing is evidenced across the studied pressure domain whereas a significant TO-mixing is achieved only near the resonance. This relates to the strength of coupling, being one order of magnitude larger in the electric case (LO) than in the mechanic one (TO). Detail is given Sec. SII.3c.

**Conclusion.** Summarizing, the mechanic properties of zincblende-Cd$_{1-x}$Zn$_x$Te are studied at the macro- ($B_0$,LO) and nano-TO scales by combining HP-RS and HP-XRD, with *ab initio* calculations in support. A regular/linear $B_0$ vs. $x$ trend is evidenced by HP-XRD. At the bond scale, the intricate inter-bond LO-electric ($x\sim0.1$) / intra-bond Zn-Te TO-mechanic ($x\sim0.9$) signals decouple / couple under pressure. Dramatic impacts on the HP-RS spectra, echoed in *ab initio* data (TO case), are decrypted using *ad hoc* versions of the Raman cross section derived within a linear dielectric approach. The net TO ($x\sim0.9$) and LO ($x\sim0.9$) coupling efficiencies, impacting the degree of mode-mixing, are compared within the same model of two coupled mechanic (TO) or electric (LO) harmonic oscillators.

Notably, a free-coupling scenario is evidenced for the two percolation-type subTO's due to the matrix-like bond, *i.e.*, self-connected in bulk, when forced to proximity by pressure. This opposes to the hindered-coupling scenario manifested earlier[7,10,11] through the achievement of a phonon exceptional point for a dispersed bond, *i.e.*, self-connected in a chain. The "alien-bond" percolation threshold emerges as the pivotal composition separating the free- and hindered-coupling regimes, supported by *ab initio* data. On the fundamental side, this endorses in a new light our view that the lattice dynamics of a disordered alloy basically falls into the scope of percolation. On the practical side, the free/hindered-coupling scenarios reveal pressure-tunable on-off phonon switches depending on the bond topology in complex media, yet unexploited.

**SUPPLEMENTAL MATERIAL**

The raw-underlying HP-XRD and HP-RS data needed to elaborate those reported in the main text, together with related supportive *ab initio* calculations and further analytical modeling of the TO-mechanic coupling are reported as supplementary material.

**ACKNOWLEDGEMENTS**

We acknowledge assistance from the PSICHÉ beamline staff of synchrotron SOLEIL (Proposal 20210410 – BAG for PSICHÉ beamline led by Y.L.G., sub-project-HP-CdZnTe co-led by A.P. & O.P.), from the IJL core facility (Université de Lorraine – http://ijl.univ-lorraine.fr/recherche/centres-de-competences/rayons-x-et-spectroscopie-moessbauer-x-gamma) for the X-ray diffraction measurements, and from P. Franchetti for the Raman measurements. VJBT in charge with the *ab initio* AIMPRO calculations acknowledge the FCT through projects LA/P/0037/2020, UIDB/50025/2020 and UIDP/50025/2020.

**AUTHOR DECLARATION**
**Conflict of Interest**
The authors have no conflicts to disclose.

**AUTHOR CONTRIBUTIONS**
**T. Alhaddad:** Data curation; Formal analysis; Investigation; Software; Validation; Visualization. **M. B. Shoker:** Data curation; Investigation. **O. Pagès:** Conceptualization; Funding acquisition (lead); Methodology; Project administration (lead); Supervision (lead); Writing original draft (lead). **A. Polian:** Investigation; Project administration (supporting); Supervision (supporting). **V. J. B. Torres:** Formal analysis, Funding acquisition (supporting); Software. **Y. Le Godec:** Resources; Project administration (supporting). **J.-P. Itié:** Investigation; Project administration (supporting). **C. Bellin:** Investigation. **K. Béneut:** Investigation. **S. Diliberto:** Investigation. **S. Michel:** Investigation. **A. Marasek:** Resources. **K. Strzałkowski:** Resources; Project administration (supporting).

**DATA AVAILABILITY**
The data that support the findings of this study are available from the corresponding author upon reasonable request.

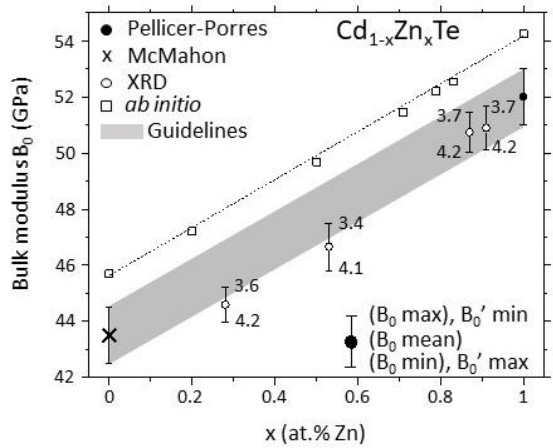

**FIG. 1: Cd$_{1-x}$Zn$_x$Te bulk-macroscopic mechanic properties.** $B_0$ values of Cd$_{1-x}$Zn$_x$Te measured in the native zincblende phase by HP-XRD (hollow circles). The experimental alloy data are surrounded by error bars corresponding to limited acceptable $B_0'$ values on fitting the pressure dependence of the unit cell volume to the Birch-Murnaghan's equation of state. Existing HP-XRD parent values (filled circles – Refs.[31,32]) are added for reference purpose. Corresponding *ab initio* (AIMPRO) data obtained on large (216-atom) zincblende-supercells optimized to a random substitution (squares) are juxtaposed, for comparison. A straight dotted line / grey area (framed by the parent error bars) is superimposed on *ab initio* /experimental data as a guideline for eye.

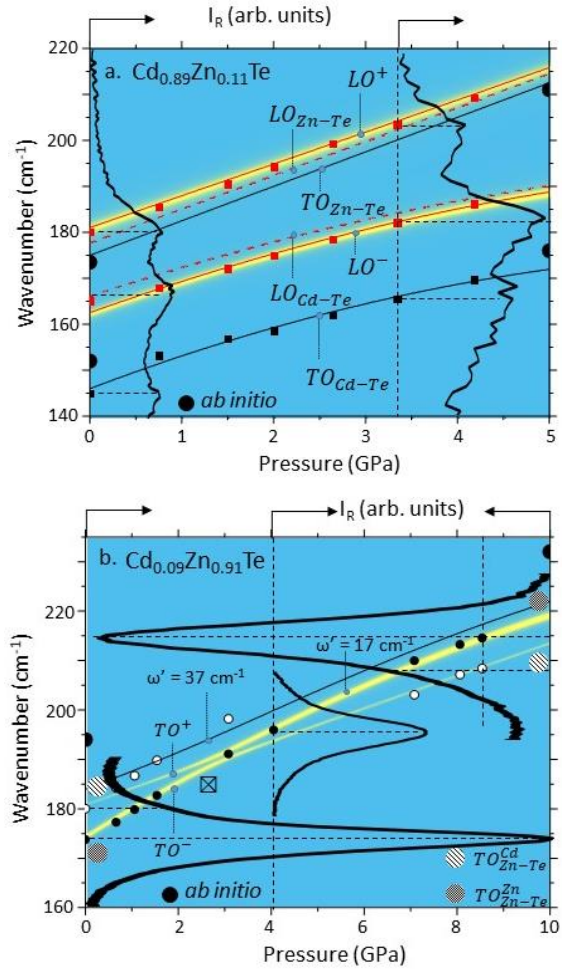

**FIG. 2: Cd$_{1-x}$Zn$_x$Te bond-microscopic mechanic properties.** Selected high-pressure LO (**a**, $x$=0.11) and TO (**b**, $x$=0.91) Cd$_{1-x}$Zn$_x$Te Raman spectra (tilted-90°) taken in the upstroke in the zincblende phase. Fair contour modeling of the Raman frequencies (enlighted curves) and intensities (thickness of curves) of the diverging-LO (squares) and converging-TO (circles) coupled features is achieved using relevant variants of the Raman cross section derived within the linear dielectric approach. In panel (**a**) the uncoupled LO's (dotted lines, not observable) and native TO's (plain lines, partly observed – filled circles) behind the (observed) coupled-LO's are indicated. The mechanical coupling is doubled near the resonance (as indicated) to account for the large repulsion of the minor (hollow circles) and main (filled circles) therein.

# Multiscale insight into the $Cd_{1-x}Zn_xTe$ vibrational-mechanical properties – High-pressure experiments and *ab initio* calculations

T. Alhaddad,[1] M.B. Shoker,[2,a)] O. Pagès,[1,b)] A. Polian,[3] V.J.B. Torres,[4] Y. Le Godec,[5] J.-P. Itié,[5] C. Bellin,[3] K Béneut,[3] S. Diliberto,[6] S. Michel,[6] A. Marasek,[7] and K. Strzałkowski[7]

[1] Université de Lorraine, LCP-A2MC, ER 4632, F-57000 Metz, France
[2] University of Luxembourg, Department of physics and materials science, 41 rue du Brill, 4422 Belvaux, Luxembourg
[3] Institut de Minéralogie, de Physique des Matériaux et de Cosmochimie, Sorbonne Université — UMR CNRS 7590, F-75005 Paris, France
[4] Departamento de Fisica and I3N, Universidade de Aveiro, 3810 – 193 Aveiro, Portugal
[5] Synchrotron SOLEIL, L'Orme des Merisiers Saint-Aubin, BP 48 F-91192 Gif-sur-Yvette Cedex, France
[6] Institute of Physics, Faculty of Physics, Astronomy and Informatics, Nicolaus Copernicus University in Toruń, ul. Grudziądzka 5, 87-100 Toruń, Poland

## Supplementary material

This annex reports on additional experimental, *ab initio* and theoretical data, besides those provided in the main text, further supporting the discussion of the $Cd_{1-x}Zn_xTe$ lattice structure / dynamics depending on pressure across the composition domain, our concern in this work. In **Sec. SI** (prefix S stands for supplementary material) the focus is on the lattice structure addressed experimentally by high-pressure X-ray diffraction (HP-XRD) at the PSICHÉ beamline of SOLEIL synchrotron. Representative raw HP-XRDiffractograms at large and small Cd contents, under focus in the main text, notably inform on the variations of lattice parameters versus pressure, from which is issued the bulk modulus at ambient pressure $B_0$ of the native zincblende phase (of central interest in this work), via a fitting of the pressure ($P$) dependence of the unit cell volume using the Birch-Murnaghan equation of state. This provides an insight into the $Cd_{1-x}Zn_xTe$ mechanic properties at the macroscopic scale. A complementary insight into the $Cd_{1-x}Zn_xTe$ mechanic properties at the bond-microscopic scale in gained in **Sec. SII** by addressing the lattice dynamics via high-pressure Raman scattering (HP-RS). Experimental (TO, LO), and *ab initio* (TO) data, reported in **Secs. SII.1** and **SII.2**, respectively, are jointly treated in **Sec. SII.3** under the same model of two coupled harmonic (mass + restoring force) oscillators, for the sake of consistency, rolled out through suitable TO (**Sec. SII.3.a**) and LO (**Sec. SII.3.b**) variants depending on the mechanic or electric nature of the restoring force, correspondingly. **Sec. SII.3** mainly elaborates on the used theoretical approaches in the main text.

A convenient preamble to this Supplementary Material section dedicated to high-pressure data is offered by Fig. S1, showing a photograph of the $Cd_{0.89}Zn_{0.11}Te$ and $Cd_{0.89}Be_{0.11}Te$ samples stowed side-by-side inside the same pressure chamber of the DAC used for their joint HP-RS studies (**Sec. SII**). All reported experimental data in this annex, regardless of type (HP-XRD or HP-RS), were obtained using a similar DAC setup (if not necessarily the same pressure transmitting medium and pressure calibrator – detail is given in the course of the discussion).

---

a) This research was performed while M.B. Shoker was at Laboratory LCP-A2MC – as specified via affiliation 1.
b) Author to whom correspondence should be addressed: olivier.pages@univ-lorraine.fr

**I. High-pressure X-ray diffraction**

I.1. Experiment

A selection of raw $Cd_{1-x}Zn_xTe$ HP-XRDiffractograms obtained at small-moderate ($x$=0.28) and large ($x$=0.89) Zn contents in the upstroke (pressure increase) at the PSICHÉ beamline of SOLEIL synchrotron using the 0.3738 Å radiation is displayed in Figs. S2a and S2b, respectively. Sharp peaks are observed at any composition and pressure, the sign of a high crystal quality. The individual peaks are labeled according to their (hkl) Miller's indices in the relevant crystal phases, abbreviated zb (zincblende, cubic), cb (cinnabar, hexagonal), rs (rock-salt, cubic) and cm (Cmcm, orthorhombic). The selected diffractograms emphasize the sensitive pressure domains corresponding to pure crystal phases and to phase coexistence.

I.2. Bulk modulus

Figs. S3a and S3b display the $P$-dependence of lattice parameters *per* crystal phase as directly inferred from the complete series of $Cd_{0.72}Zn_{0.28}Te$ and $Cd_{0.11}Zn_{0.89}Te$ HP-XRDiffractograms, partly reported in Fig. S2a and S2b. The corresponding unit cell volume vs. $P$ variations in the native zincblende phases, shown in Fig. S4a, are fitted to the Birch-Murnaghan equation of state (curves).[18] The $B_0$ and $B'_0$ values resulting from the fit are reported in Fig. 1. Generally, $B'_0$ pre-fixes the slope of the fitted curve and then $B_0$ comes out as a by-product of the fit. The only admitted $B'_0$ values are such that the fitted curve passes through all available experimental data within error bars. The as-found $B'_0$ values repeatedly fall within the range 3.5 – 4.2 at any studied composition $x$ (Fig. 1). This falls close to the ZnTe value, fixed to 4 in Ref.,[31] and not far from the CdTe one, estimated at 6.4±0.6, Ref.[32] Within error bars the fitted $B_0$ values (Fig. 1, hollow circles) linearly increase with $x$ (dotted line) between the CdTe (42±2 GPa)[32] and ZnTe (52±2 GPa)[31] values also determined by HP-XRD, consistently with *ab initio* predictions (squares).

Another parameter coming into the fit is the reference unit cell volume at 0 GPa. This fixed to the value derived from powder XRD measurements done at ambient conditions using the Cu K$\alpha$ line. Such diffractograms (examples are available in Ref.[3]) reveal a linear dependence of the (unique) lattice constant $a$ vs. the composition $x$, as independently determined by chemical analysis,[3] across our series of free-standing zincblende-type $Cd_{1-x}Zn_xTe$ single crystals (Fig. S4b, filled circles). The linear $a$ vs. $x$ trend is replicated in current *ab initio* data (hollow circles) and is further consistent with the Vegard' law obeyed by the lattice constant of similar $Cd_{1-x}Zn_xTe$ single crystals whose composition was determined via the Cd:Zn ratio by wavelength-dispersive X-ray spectroscopy.[S1]

**II. High-pressure Raman scattering**

II.1. Experiment

HP-RS experiments are performed at symmetric compositions at both ends of the $Cd_{1-x}Zn_xTe$ composition domain, *i.e.*, with $Cd_{0.89}Zn_{0.11}Te$, on the one hand, in search for the LO modes, and with $Cd_{0.11}Zn_{0.89}Te$, on the other hand, focusing on the TO modes (see main text).

The LO insight into $Cd_{0.89}Zn_{0.11}Te$ is achieved in near resonant conditions with the 488.0 nm operated at 77K. A reference $Cd_{0.89}Be_{0.11}Te$ sample placed in the same pressure chamber of the DAC (Fig. S1) is used for *in situ* calibration of the available Cd-Te oscillator strength across the studied pressure domain (see below). Somewhat fortuitously, the two samples have exactly the same composition within a half percent, so that the amount of Cd-Te oscillator strength found with $Cd_{0.89}Be_{0.11}Te$ can be re-used as such for $Cd_{0.89}Zn_{0.11}Te$ (being clear that a difference in alloy composition would not have been problematic, anyway). Another fortuitous analogy between $Cd_{0.89}Be_{0.11}Te$ and $Cd_{0.89}Zn_{0.11}Te$ is that they both transite from zincblende to rock-salt at the same critical pressure, *i.e.*, ~5.5 GPa (within experimental resolution), referring to existing HP-XRD data in the literature at nearly ($Cd_{0.93}Be_{0.07}Te$ – Ref.[11]) or exactly the same ($Cd_{0.89}Zn_{0.11}Te$ – Ref.[29]) compositions.

The LO-like Raman spectra jointly obtained with $Cd_{0.89}Zn_{0.11}Te$ (thick lines) and $Cd_{0.89}Be_{0.11}Te$ (thin curves) in the upstroke across the pressure domain of their native zincblende phase are reported in

Fig. S5. The Be-Te bond is most covalent-stiff/light among II-VI's,[30] generating an impurity mode far away, *i.e.*, around 400 cm$^{-1}$, from the Raman modes due to the comparatively ionic-soft/ heavy Cd-Te and Zn-Te matrix-like bonds, showing up around 175 cm$^{-1}$. Hence, the $LO_{Cd-Te}$ and $LO_{Be-Te}$ modes of Cd$_{0.89}$Be$_{0.11}$Te are well-separated, hence decoupled. As such, they are assigned to specific bonds (as indicated by the subscript). In contrast, the coupled $LO^-$ and $LO^+$ modes of Cd$_{0.89}$Zn$_{0.11}$Te, ranked in order of increasing frequency (refer to the superscript), vibrate nearby.[2] The $TO_{Cd-Te}$ mode shows up distinctly at any pressure with both alloys. As for $TO_{Be-Te}$, it is quasi degenerated with $LO_{Be-Te}$ at the considered minor Be content.[3]

Unpolarized Raman spectra taken with the 632.8 nm laser line at 300 K across the native zincblende phase of Cd$_{0.11}$Zn$_{0.89}$Te and at extreme pressures of the native zincblende phase of Cd$_{0.13}$Zn$_{0.88}$Te (0−9 GPa) in the upstroke (plain lines) and in the downstroke (dotted line) are shown in Figs. 6a and 6b, respectively. At both ends of the pressure domain, the TO modes are quasi decoupled (see main text), and hence labeled with a subscript and a superscript specifying the bond vibration and the local environment ("same" or "alien"). The ZnTe-like TO's are forced to proximity by pressure and do mechanically couple, rendering the above bond / environment-specific assignment obsolete. A neutral notation is then used in terms of $TO^-$ and $TO^+$, in order of increasing frequency. Simplicity is likewise retained for labeling, on the one hand, the main LO feature due to the dominant Zn-Te bond, noted $LO_{Zn-Te}$, and, on the other hand, the CdTe-like impurity mode with degenerate TO-LO character at the studied minor Cd content – identified as $O_{Cd-Te}$.

II.2. *Ab initio* calculations

Fig. S7 reports on *ab initio* (AIMPRO) TO-like Raman spectra calculated at ambient (thick curves) and high (5 GPa, thin curves) pressures using large (216-atom) disordered zincblende-Cd$_{1-x}$Zn$_x$Te supercells with minor-to-moderate Zn contents ($x$=0.83, 0.79, 0.7 and 0.5) on both sides of the Cd-Te bond percolation threshold ($x_{Cd-Te}$=0.81).[2] With increasing Zn content the high-frequency ZnTe-like signal strengthens at the expense of the low-frequency CdTe-like one, as expected. At small Cd content ($x \geq 0.79$) a minor feature emerges at 5 GPa on the low-frequency tail of the main ZnTe-like band, that was absent at 0 GPa. The minor feature disappears on moving away from the Cd-Te bond percolation threshold ($x$~0.3, 0.5). By analogy with the reported experimental TO-Raman data at large Zn content (Fig. S6), the minor and dominant features showing up at 5 GPa are assigned in terms of $TO_{Zn-Te}^{Zn}$ and $TO_{Zn-Te}^{Cd}$, respectively. Accordingly, at minor Cd content, the ZnTe-like doublet would suffer an inversion between 0 and 5 GPa, suggesting the free mechanic coupling of oscillators when forced to proximity by pressure. Oppositely, the absence of minor feature at large Cd content, suggests the achievement of a phonon exceptional point, corresponding to a hindered mechanic coupling on crossing at the resonance.[7] Altogether, this nicely recollects with the picture outlined in the main text, emphasizing $x_{Cd-Te}$ as a pivotal composition separating the free- and hindered-coupling regimes of the Zn-Te subTO's of Cd$_{1-x}$Zn$_x$Te.

II.3. Modeling – linear dielectric approach

A sensitive input parameter coming into our analytical expression of the Raman cross section (RCS) used for contour modeling of the uncoupled-TO and LO Raman lineshapes, given by Eq. 1 of Ref.[3], or its variant for coupled-TO, given by Eq. 7 of Ref.[10], is the amount of oscillator strength awarded to a given mode at a given pressure. In an alloy, this scales as the fraction of the corresponding oscillator in the crystal.[6] Once the latter is properly determined, as recently achieved within the PM-version worked out for Cd$_{1-x}$Zn$_x$Te,[3] it all comes down to elucidate the pressure dependence of the parent (CdTe and ZnTe) oscillator strengths. These express as $S = \varepsilon_\infty \cdot \frac{\omega_L^2 - \omega_T^2}{\omega_T^2}$, with ($\omega_T$, $\omega_L$) standing for the (TO, LO) Raman frequencies and $\varepsilon_\infty$ representing the high-frequency dielectric constant (at $\omega \gg \omega_T$).[6] In a first approximation, $\varepsilon_\infty$ is taken constant for ZnTe$^{S2}$ and CdTe$^{S3}$, based on existing calculations using the *ab initio* WIEN2k code and a semi-empirical tight-binding method, respectively. These indicate for ZnTe a stability of $\varepsilon_\infty$ within 4% between 0 and 20 GPa and for CdTe a variation of $\varepsilon_\infty$ less than 9% between 0 and 5 GPa, matching the pressure domain probed in the current Cd$_{0.89}$Zn$_{0.11}$Te HP-RS study.

As for the pressure dependence of the $\omega_T - \omega_L$ optic band, this is well-documented for ZnTe.[28] For CdTe, only the pressure dependence of $\omega_L$ could be accessed experimentally.[S4]

II.3.a. $Cd_{1-x}Zn_xTe$ LO Raman signal ($x$~0.1)

An access to the CdTe oscillator strength, $S_{CdTe}$, is especially important at large Cd content where the HP-RS LO study is carried out. A direct insight depending on pressure for $Cd_{0.89}Zn_{0.11}Te$ is currently achieved *in situ* by modeling, using the RCS given by Eq. (1) in Ref.[3], the TO-LO splitting manifested by the uncoupled Cd-Te mode of $Cd_{0.89}Be_{0.11}Te$, used as an internal calibrator, stowed side-by-side with $Cd_{0.89}Zn_{0.11}Te$ inside the pressure chamber of the DAC (Fig. S1). Besides $\varepsilon_\infty$, the only used input parameter is the $TO_{Cd-Te}$ Raman frequency, taken from experiment (Fig. S5). $S_{CdTe}$ is adjusted until the LO version of the $Cd_{0.89}Be_{0.11}Te$-RCS generates a $LO_{Cd-Te}$ mode exactly at the observed frequency (Fig. S5). $S_{CdTe}$ is then re-injected as such in the $Cd_{0.89}Zn_{0.11}Te$ RCS to generate blindly the $LO^-$ and $LO^+$ coupled features at the same pressure. Again, the main input parameters are the TO frequencies, *i.e.*, of the matrix-like $TO_{Cd-Te}$ and of the impurity-like $TO_{Zn-Te}$ ($\omega_{imp}$). The former is visible in experiment (Fig. S5, filled squares in Fig. 2a), but not the latter. In first approximation, a linear $\omega_{imp}$ vs. $P$ variation is assumed. The starting value at 0 GPa is that reported in Ref.[3] for the $TO_{Zn-Te}^{Cd}$ submode of $Cd_{0.89}Zn_{0.11}Te$ at 0 GPa. The searched slope is then adjusted so that the theoretical $LO^-$ and $LO^+$ frequencies (emphasized in Fig. 2a) best match the experimental values (hollow symbols in Fig. 2a).

The optimized system of uncoupled TO frequencies (plain curves), coupled LO frequencies (emphasized – Fig. 2a or thick – Fig. S5 curves) and underlying uncoupled LO frequencies (dashed curves) is depicted *in extenso* in Fig. 2a. *Ab initio* TO frequencies calculated with a CdTe-like (matrix-like) supercell containing an isolated (impurity) duo of Zn atoms (connected via Te) at extreme pressures of the studied domain (0 − 5 GPa), remarkably consistent with experiment for both the matrix- and impurity-related modes, are added for comparison (filled circles). Corresponding Raman lineshapes are shifted beneath experimental signals at selected pressures in Fig. S5, for a direct insight. The modeling of the raw-uncoupled LO Raman signal due to a given bond is achieved by omitting the phonon contribution due to the undesired oscillator in the alloy-related relative dielectric function coming into the resonance term of the standard LO version of the RCS – as detailed in Ref.[3].

Alternative input parameters besides the parent oscillator strength coming into the expression of the RCS used to model the high-pressure MREI-like LO Raman signal at 77 K, are taken from the MREI data set identified by Talwar *et al.*[33] Additional input parameters are the Faust-Henry coefficients of the parent CdTe (-0.14) and ZnTe (-0.11) compounds,[3] assumed to be independent of pressure in a crude approximation.

II.3.b. $Cd_{1-x}Zn_xTe$ TO Raman signal ($x$~0.9)

Though they refer to polar cation-anion bonds, the TO modes detected in a conventional backscattering Raman experiment, as in the present case, are deprived of electric field. This is because the transferred wavevector, maximum in such experiment, is not compatible with the propagation of a photon-like (transverse) electric field.[S5] Such purely mechanic TO's hardly couple, even when they vibrate nearby.[S6] The corresponding RCS is the TO version of the given expression in Eq. (1) of Ref.[3]

Now, some mechanic coupling has to be taken into account when neighboring TO's are forced to extreme proximity and cross by application of some relevant external stimulus, as currently achieved between the two Zn-Te subTO's of $Cd_{0.11}Zn_{0.89}Te$ under pressure. As already mentioned, there are two sub-scenarios at the resonance/crossing, *i.e.*, either the free-coupling or the hindered-coupling. The hindered-coupling manifesting the achievement of a so-called phonon exceptional point at the resonance was modeled by adapting a model of two overdamped mechanically-coupled harmonic oscillators originally developed by Dolfo and Vigué,[57] applied to $Zn_{0.5}Be_{0.5}Se$ as a case study.[7] The TO Raman cross section in the free-coupling case was lately derived for undamped harmonic oscillators, given by Eq. (7) of Ref.[10].

For contour modeling of the ZnTe-like ($TO^-$,$TO^+$) Raman doublet of $Cd_{0.11}Zn_{0.89}Te$ in its pressure dependence, displayed in Fig. 2b, we use a simplified form of the latter RCS in which the coupling-term, preserved at the denominator, is disregarded with respect to the raw-uncoupled terms at the numerator. The $S_{ZnTe}$ oscillator strength is involved, whose $P$-dependence is well-documented in the literature, as already mentioned (Sec. II.3). Alternative input parameters are the $P$-dependencies of the TO frequencies of the two underlying raw-uncoupled modes. For the main $TO_{Zn-Te}^{Zn}$ we use the variation carefully measured for the TO mode of ZnTe by Camacho *et al.*,[28] downshifted to the observed $TO_{Zn-Te}^{Zn}$ value for Cd$_{0.11}$Zn$_{0.89}$Te at 0 GPa. For the minor $TO_{Zn-Te}^{Zn}$, we make an assumption, justified *a posteriori*, that the weak mechanic coupling does not challenge its frequency away from the resonance, identified at ~3 GPa (symbolized ⊠ in Fig. 2b). A linear variation is considered between the extreme values at both ends of the studied pressure domain. A minimal phonon damping (1 cm$^{-1}$) is uniformly taken, for a clear separation of close ($TO^-$, $TO^+$) features at any pressure, notably close to the resonance.

II.3.c. TO vs. LO pressure-induced coupling processes

In the main text we compare the $P$-induced TO-coupling (Cd$_{0.11}$Zn$_{0.89}$Te) and LO-decoupling (Cd$_{0.89}$Zn$_{0.11}$Te) processes using the same model of two undamped harmonic oscillators, for the sake of consistency, either coupled mechanically (TO) or electrically (LO). The corresponding generic system of mechanic / polarization equations, worked out in line with earlier work done on the Be-Se doublet of Zn$_{0.5}$Be$_{0.5}$Se[7] is given by (4) in Ref.[10] Presently, the focus is on the dynamic matrix $\widetilde{M}_{T,L}$ (with subscript specifying the transverse or longitudinal symmetry). Hence, the generalized forces used to derive the Raman cross section are omitted.

By substituting $Q_{i,T} = \sqrt{\mu_i} u_i$ in the TO case (subscript), and $Q_{i,L} = \sqrt{x_i \mu_i} u_i$ in the LO case, for the actual $i$-bond stretching $u_i$ ($i$=1,2), characterized by its reduced mass $\mu_i$ and bond fraction $x_i$, the dynamical matrices $\widetilde{M}_{T,L}$ turn symmetric. The off-diagonal term $V$ represents the strength of mechanic/electric coupling. The difference $2\Delta = |E_1 - E_2|$ between diagonal terms marks the proximity to the resonance. The unit $|Q_\pm\rangle$ wavevectors are orthogonal, following from, *e.g.*, $|Q_\pm\rangle = \begin{pmatrix} \cos\theta \\ \sin\theta \end{pmatrix}$. Accordingly, $\cos^2\theta$ and $\sin^2\theta$ specify how much, in percent, the raw-uncoupled $|Q_1\rangle$ and $|Q_2\rangle$ oscillators are involved in $|Q_+\rangle$, respectively, and vice versa for $|Q_-\rangle$. The net coupling efficiency is captured via $|tan2\theta|=V/\Delta$, *i.e.*, in terms of a compromise. The $P$-dependence of the cited parameters in the TO and LO cases are compared in Fig. 2c. $|Tan2\theta|$ achieves maximum on crossing the TO resonance at ~4 GPa, without LO parallel since $LO^-$ and $LO^+$ break away under pressure. Yet, a significant LO-mixing, attested the $\cos^2\theta$ and $\sin^2\theta$ values being comparable, persists across the pressure domain. In contrast, the TO's mix significantly only near the resonance. This is because $V_T \ll V_L$ (refer to arrows – Fig. S8a), justifying *a posteriori* our approximation to omit the mechanic coupling in the LO case.

In the LO case, the oscillators couple both electrically, via a Coulomb restoring force involving their common macroscopic electric field $E$ due to the bond ionicity, and also mechanically, as the TO modes deprived of electric field (Sec. SII.3b). The mechanic coupling is neglected in this case, in a crude approximation (see main text). The relative oscillator fractions are materialized in the dynamic matrix $\widetilde{M}_L$ via the polarization equation, *i.e.*, $\varepsilon_r=0$ for a LO mode.[55] Such materialization disappears with the purely-mechanic ($E$=0) TO Raman mode probed in the conventional backscattering geometry,[55] used in this work. Strictly, such TO approach makes sense only if both oscillators are equally represented in the crystal, as in the Zn$_{0.5}$Be$_{0.5}$Se case.[7] This is not so in the present Cd$_{0.11}$Zn$_{0.89}$Te case, in which one oscillator is dominant ($TO_{Zn-Te}^{Zn}$) and the other minor ($TO_{Zn-Te}^{Cd}$). This justifies a more refined approach to weight the mechanic coupling suffered by a given ($u_i$) oscillator by the fraction ($x_j$) of the other ($u_j$) oscillator in the crystal. By doing so, $\widetilde{M}_T$ turns asymmetric so that the net coupling efficiency ($\sqrt{V_{12} V_{21}}/\Delta$) involves the geometrical means of the off-diagonal $\widetilde{M}_T$-terms. However, this generates few changes in its pressure dependence with respect to the symmetric case (Fig. 2c), as apparent in Fig. S8b.

**Supplementary material – only references**

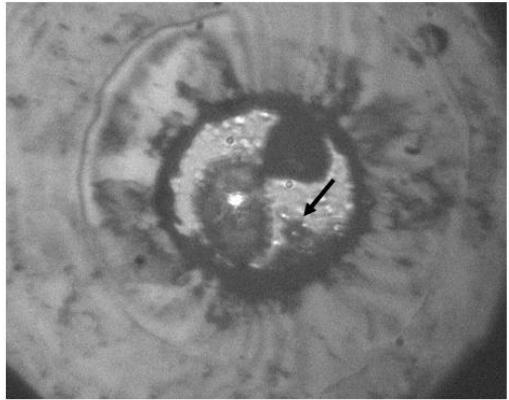

**FIG. S1**: **DAC pressure chamber for HP-RS on CdTe-like alloys.** Post-downstroke photograph of the clear/transparent $Cd_{0.89}Zn_{0.11}Te$ and dark/absorbing $Cd_{0.89}Be_{0.11}Te$ samples stowed side-by-side inside the pressure chamber of the DAC used for the HP-RS measurements, besides the ruby (pointed by an arrow) used for pressure calibration. A spotlight is generated by the laser beam at the impact spot on the sample surface.

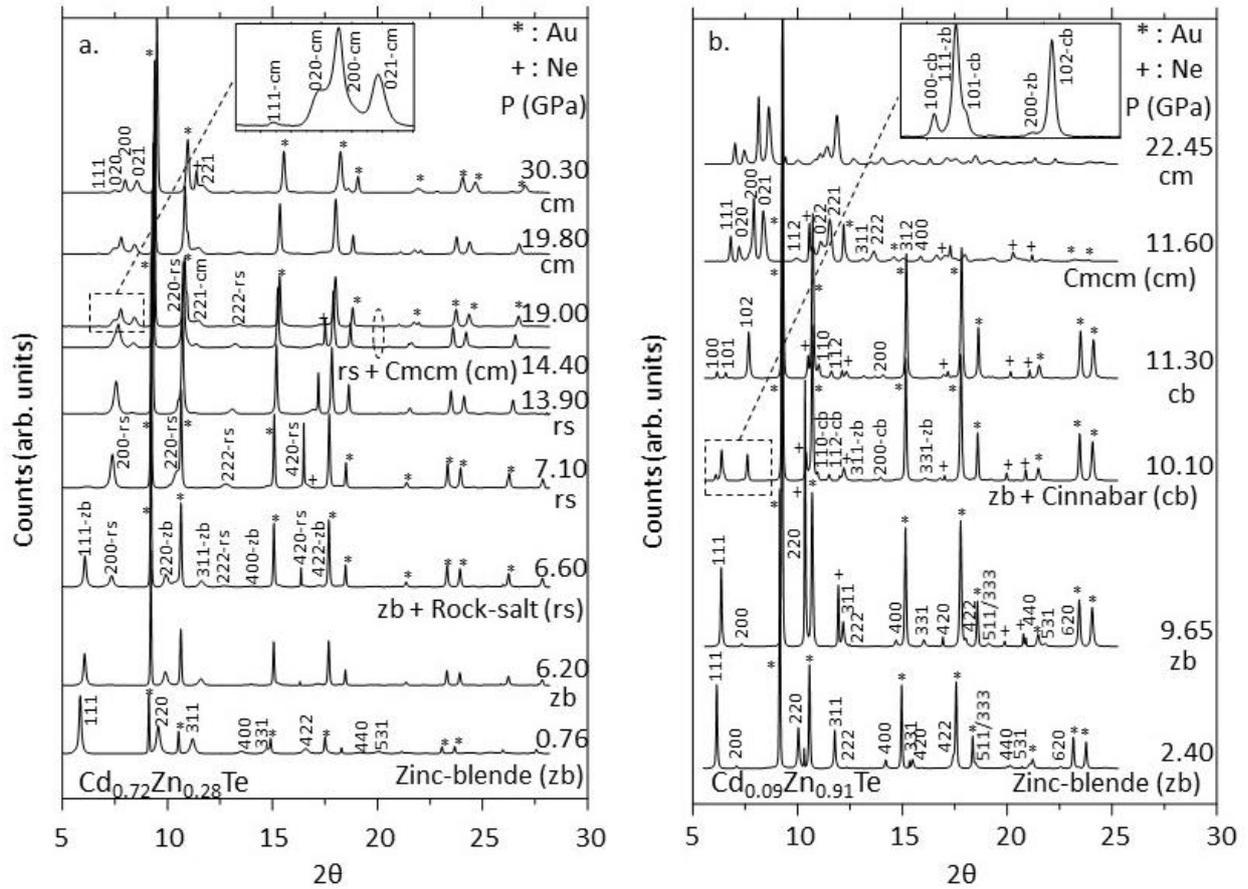

**FIG. S2**: **High-pressure $Cd_{1-x}Zn_xTe$ X-ray diffractograms.** Selected **(a)** $Cd_{0.72}Zn_{0.28}Te$ and **(b)** $Cd_{0.09}Zn_{0.91}Te$ powder HP-XRD diffractograms taken in the upstroke with special attention to the pure-phase and mixed-phase domains. The individual peaks are labelled using the (hkl) Miller indices of diffraction planes in various (zincblende-zb, cinnabar-cb, rocksalt-rs, Cmcm-cm) structural phases. Additional diffraction peaks originate from Au and Ne used for pressure calibration and as the pressure transmitting medium, respectively, as indicated.

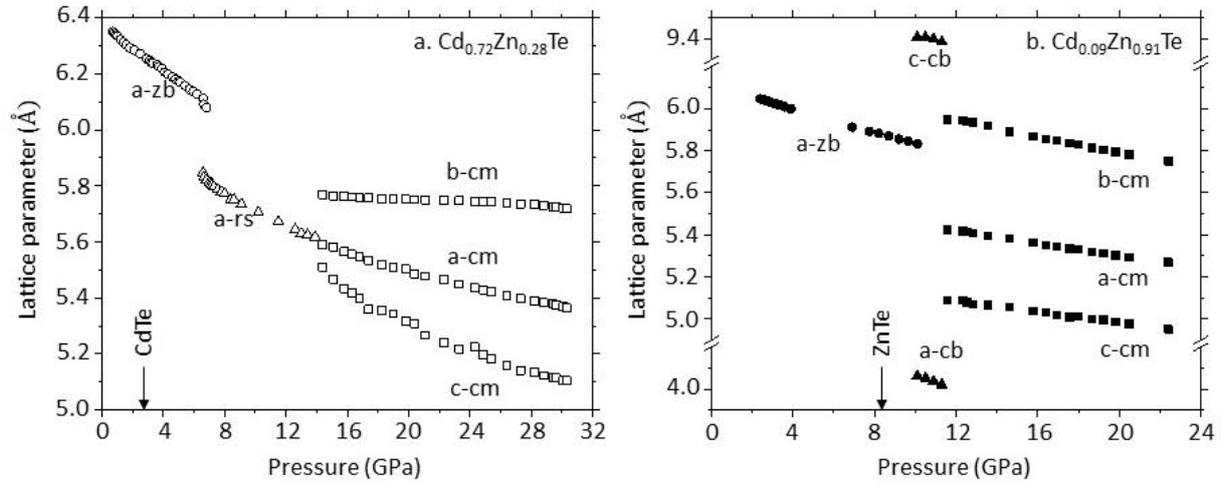

**Fig. S3**: **Pressure dependence of $Cd_{1-x}Zn_xTe$ lattice constants**. Pressure dependence of the **(a)** $Cd_{0.72}Zn_{0.28}Te$ and **(b)** $Cd_{0.09}Zn_{0.91}Te$ lattice constants measured by powder high-pressure X-ray diffraction (**Fig. S2**) across successive structural phases (zincblende-zb, cinnabar-cb, rocksalt-rs, Cmcm-cm) in the upstroke. The critical pressures corresponding to first departure from the native zincblende phase of the immediately related pure compounds are indicated (arrows), for reference purpose.

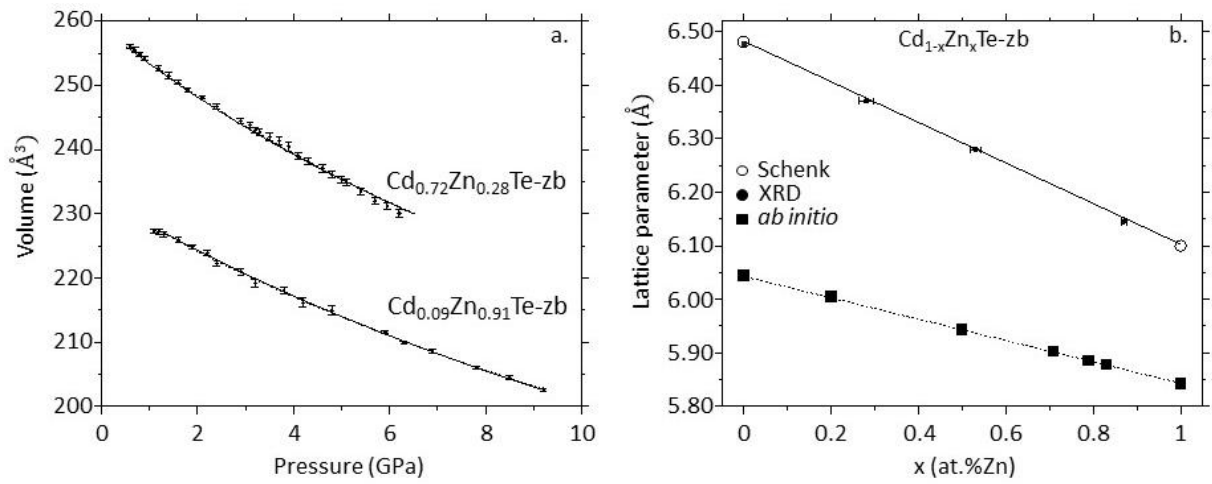

**Fig. S4**: **Derivation of the $Cd_{1-x}Zn_x$Te bulk modulus. (a)** $Cd_{1-x}Zn_x$Te (x=0.28, 0.89) unit cell volume vs. $P$ dependence of the native zincblende (zb) phase measured by powder HP-XRD fitted to the Birch-Murnaghan equation of state (curves). The resulting $B_0$ and $B_0'$ values are reported in Fig. 1. **(b)** Underlying $x$-dependence of the lattice constant measured by powder XRD at ambient conditions (circles, marred by error bars). A linear trend (plain line) between existing parent values, evidenced in the literature,[S1] is replicated in *ab initio* data (squares, connected via a straight-dotted line as a guideline for eye).

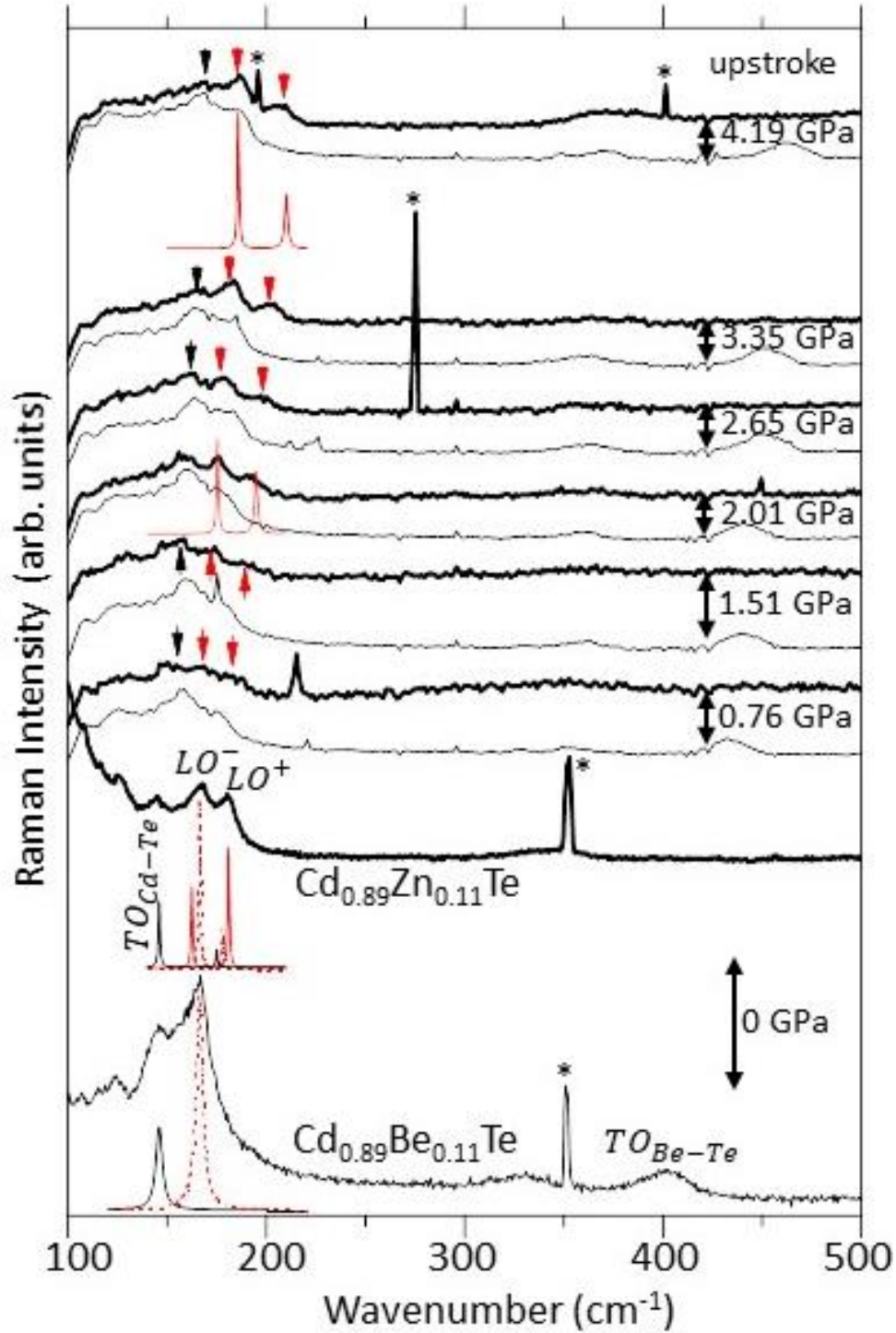

**Fig. S5**: **LO-like high-pressure $Cd_{0.89}Zn_{0.11}Te$ and $Cd_{0.89}Be_{0.11}Te$ Raman spectra**. High-pressure Raman spectra taken in the upstroke across the native zincblende phases of $Cd_{0.89}Zn_{0.11}Te$ (thick curves) and $Cd_{0.89}Be_{0.11}Te$ (thin curve) – used for *in situ* calibration of the Cd-Te oscillator strength – stowed side-by-side in the pressure chamber of the DAC (Fig. S1). Theoretical Raman signals of the coupled-LO's (plain lines), of the underlying uncoupled-LO's (dotted line) and of the native TO's (thick lines), calculated via relevant versions of the analytical Raman cross section, are shifted beneath the corresponding raw-experimental signals at well-spanned pressures, for a direct comparison. The star marks spurious signals or laser lines.

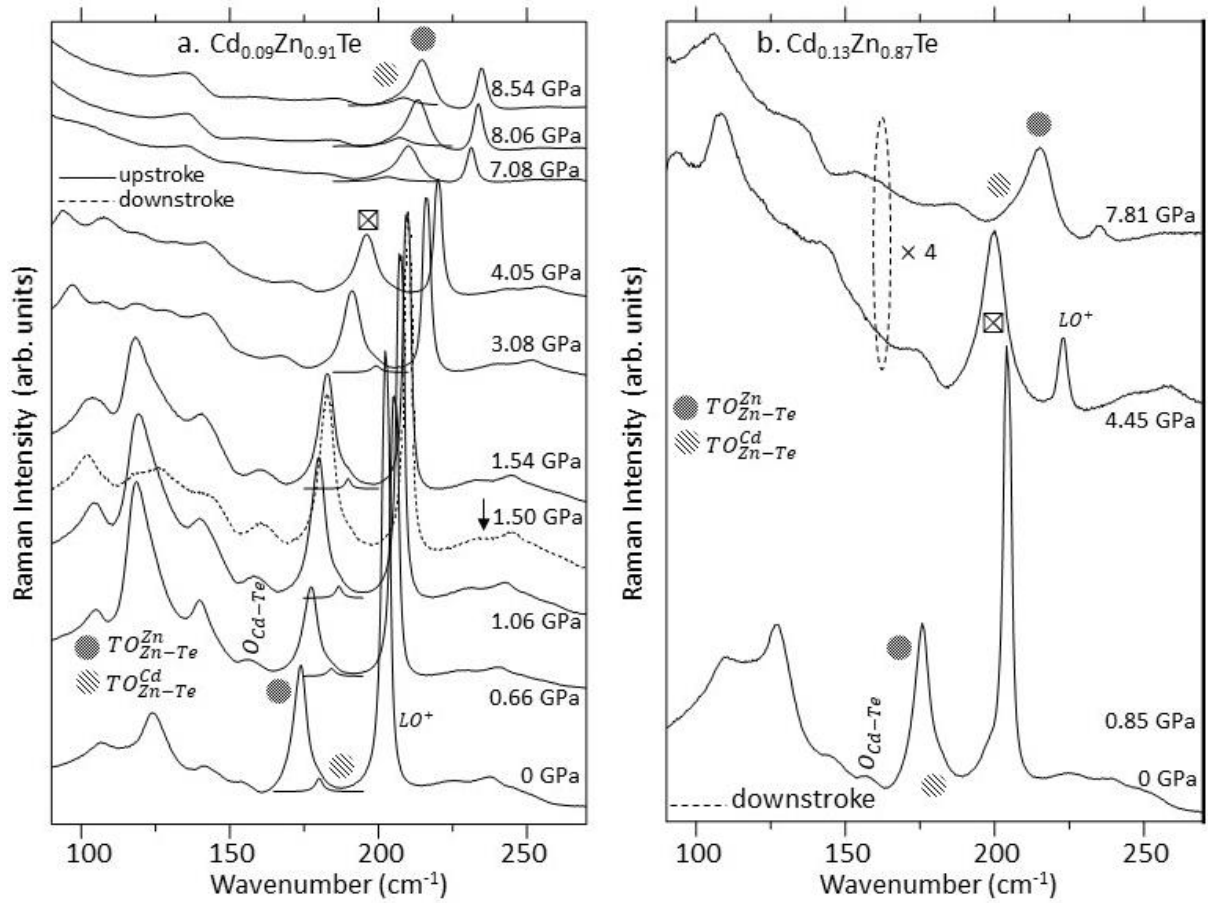

**Fig. S6**: **High-pressure Cd$_{1-x}$Zn$_x$Te Raman spectra ($x \sim 0.9$)**. High-pressure Raman spectra taken across the native zincblende phase of **(a)** Cd$_{0.09}$Zn$_{0.91}$Te and **(b)** Cd$_{0.13}$Zn$_{0.87}$Te in the upstroke (plain curves) and downstroke (dotted curves). The dominant and minor (emphasized) ZnTe-like TO features suffer an inversion under pressure (sketched out by using dotted lines) with a crossing point around 4 GPa. The corresponding Raman signal is accordingly symbolized by a cross (⊠).

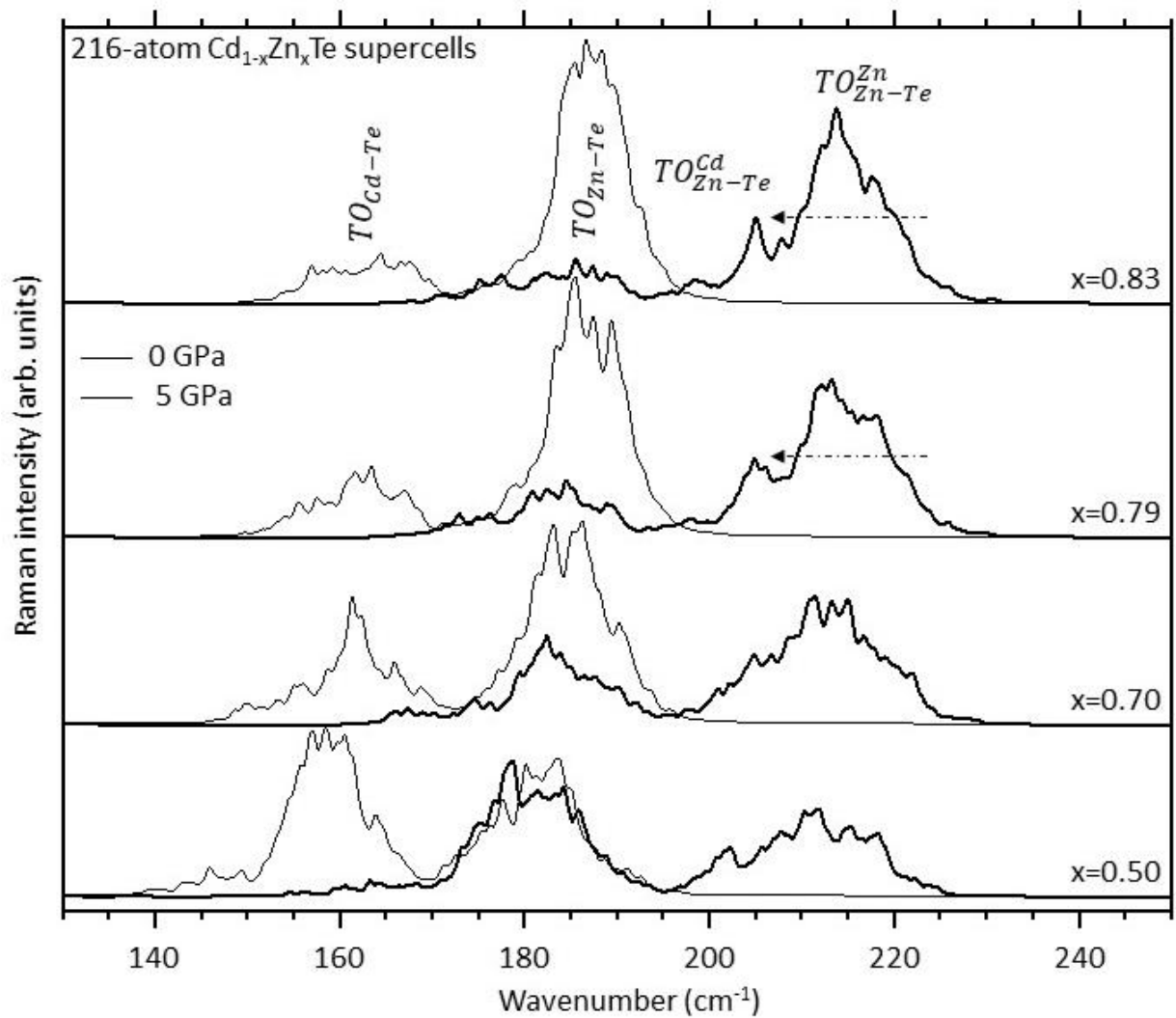

**Fig. S7**: *Ab initio* high-pressure Cd$_{1-x}$Zn$_x$Te **Raman spectra.** High-pressure TO-like Cd$_{1-x}$Zn$_x$Te Raman spectra obtained by applying the *ab initio* AIMPRO code to large (216-atom) zinblende-supercells with minor-to-intermediary Cd content optimized to a random Cd↔Zn substitution. At high pressure, a minor feature emerges distinctly on the low frequency side of the dominant feature, suggesting a TO-inversion (emphasized by arrows).

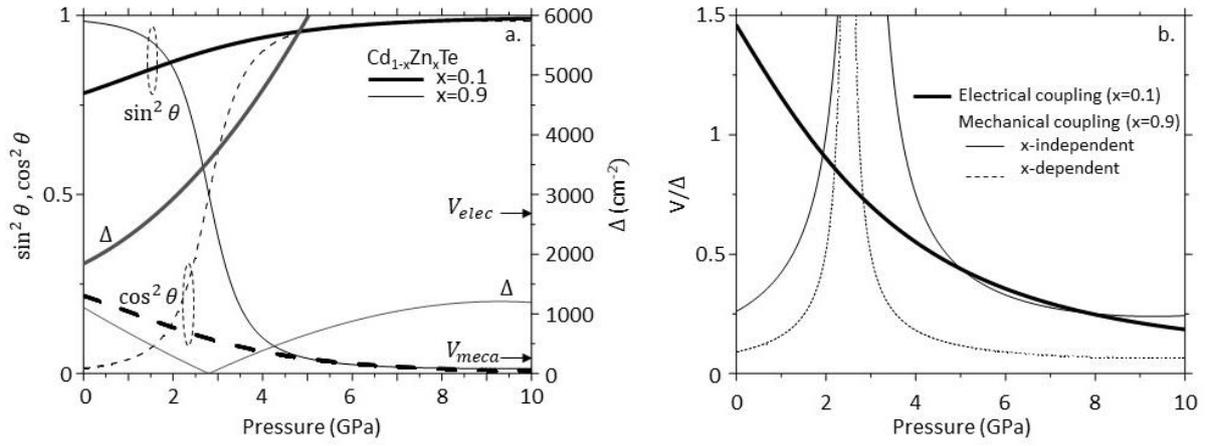

**Fig. S8**: **LO vs. TO couplings.** **(a)** TO (thin lines) and LO (thick lines) degrees of mode mixing $cos^2\theta$ and $sin^2\theta$ resulting from a compromise between $V$ (differentiated in the electric and mechanic cases) and $\Delta$ in the symmetric/$x$-independent case. **(b)** Net efficiencies $V/\Delta$ of the $Cd_{1-x}Zn_xTe$ LO (thick curve, $x$=0.11) and TO (thin curves, $x$=0.89) couplings, with a distinction between the symmetric/$x$-independent (plain line) and antisymmetric/$x$-dependent (dotted line) cases in the TO symmetry.